\documentclass[12pt]{article}
\usepackage{graphicx}
\usepackage{amssymb}
\usepackage{epstopdf}
\usepackage{multirow}
\newcommand\mathC{\mkern1mu\raise2.2pt\hbox{$\scriptscriptstyle|$}
        {\mkern-7mu\rm C}}              %%% The complex  numbers
\newcommand{\be}{\begin{equation}}
\newcommand{\ee}{\end{equation}}

\let\ssection=\section
\renewcommand{\section}{\setcounter{equation}{0}\ssection}

\begin{document}
\begin{center}
{\large\bf On Time {\em chez} Dummett}
\end{center}

\begin{center}
Jeremy Butterfield\\
Trinity College, Cambridge CB2 1TQ: jb56@cam.ac.uk
\end{center}

\begin{center}    
Published in {\em The European Journal of Analytic Philosophy} {\bf 8} (2012), 77-102; (a special issue about philosophy of physics, in honour of Michael Dummett)
 \end{center}


\begin{center} Sat 12 May 2012 \end{center}

\begin{abstract}
I discuss three connections between Dummett's writings about time and philosophical aspects %%@
of physics.

The first connection (Section 2) arises from remarks of Dummett's about the different %%@
relations of observation to time and to space. The main point is uncontroversial and %%@
applies equally to classical and quantum physics. It concerns the fact that perceptual %%@
processing is so rapid, compared with the typical time-scale on which macroscopic objects %%@
change their observable properties, that it engenders the idea of a `common now', spread %%@
across space.

The other two connections are specific to quantum theory, as interpreted along the lines of %%@
Everett. So for these two connections, the physics side is controversial, just as the %%@
philosophical side is.

In Section 3, I connect the subjective uncertainty before an Everettian `splitting' of the %%@
multiverse to Dummett's suggestion, inspired by McTaggart,  that a complete, i.e. %%@
indexical-free description of a temporal reality is impossible. And in Section 4, I connect %%@
Barbour's denial that time is real---a denial along the lines of Everett, rather than %%@
McTaggart---to Dummett's suggestion  that statements about the past are not determinately %%@
true or false, because they are not effectively decidable.

\end{abstract}
\newpage

\section{Introduction}\label{intro}
Michael Dummett was undoubtedly one of the most significant philosophers of the last fifty %%@
years. So it is a privilege to honour his memory, and his work. Given that work's emphasis %%@
on the philosophy  of language and logic, it is something of a question for me, as a %%@
specialist in philosophy of physics, how best to do so: a question made harder by the %%@
contrast between his semantic anti-realism (justificationism) and my own realism. What I %%@
propose is to take three topics in the philosophy of time which Dummett's writings have %%@
addressed, and to report on their connection to philosophy of physics. In Section %%@
\ref{tms}, I briefly review the  themes in philosophy of time that will be relevant. (I %%@
will make a point of giving many references, with a view to allaying Dummett's worry that %%@
`specialist philosophers of physics speak a technical language among themselves, and fail %%@
to communicate with other philosophers in the mainstream' (2007, 25).) 

The first topic (Section \ref{seepres}) is uncontroversial (at least, so I say!), and %%@
applies equally to classical and quantum physics. It concerns the fact that usually, %%@
perceptual processing and oral communication is so rapid, compared with the typical %%@
time-scale on which macroscopic objects change their observable properties, that it %%@
engenders the idea of a `common now', spread across space. Within Dummett's writings, the %%@
springboard for this is some remarks in his {\em Frege: the Philosophy  of Language}, about %%@
how observation relates differently to time than it does to space (1973, 388).

The other two connections are specific to quantum theory, as interpreted along the lines of %%@
Everett. So here, both the physics and philosophy sides of the connection are %%@
controversial. Thus in Section \ref{split}, I connect the subjective uncertainty before an %%@
Everettian `splitting' of the multiverse to Dummett's  suggestion that a complete, i.e. %%@
indexical-free description of a temporal reality is impossible. So far as I know, Dummett %%@
first made this suggestion in his (1960); but it is echoed later, in his 2002 Dewey %%@
lectures (2003, 2004). Dummett takes this suggestion as the moral of McTaggart's `proof' %%@
that time is unreal.

This latter, dizzying, idea leads in to my third topic. In Section \ref{julian}, I report %%@
Barbour's (1999) denial that time is real: though in another sense than McTaggart's---a %%@
sense that in effect combines the ideas  of Everett and Arthur Prior. I connect this denial %%@
to an idea which Dummett has formulated and explored: hoping, I should add, to find grounds %%@
on which he could reject it, rather than grounds for  accepting it. Roughly speaking, it is %%@
the idea that statements about the past, if true at all, are true only in virtue of present %%@
traces (including memories). So far as I know, Dummett first explored this idea in his %%@
(1969); but he returned to it in more detail (again, hoping to reject it, not embrace it) %%@
in his (2003, 2004).

I end this preamble with two preliminary comments. First: the connections I state are %%@
intellectually robust. But I should issue a health warning, about the material in Sections %%@
\ref{split} and \ref{julian}. Namely: the physical ideas, to which I there connect %%@
Dummett's writings, are highly controversial.  Of course, it is well known that the Everett %%@
interpretation (Section \ref{ustdsplit}) is controversial. But I should emphasize that %%@
Barbour's denial of time (Section \ref{julian2}) is even more controversial. In short: %%@
though I endorse these Sections' two bridges from Dummett to physics---I do not vouch for %%@
the truth of what is on the far, physics, side of them!

Second: although my philosophical temperament is, unlike Dummett's, realist and naturalist %%@
(as will be clear from Section \ref{seepres} et seq.), I would like to pay tribute at the %%@
outset to  Dummett's philosophical imagination. I especially admire his over-arching theme %%@
that empirical propositions that are not effectively decidable  should be treated along the %%@
lines intuitionists advocate for mathematical propositions: viz. as not determinately true %%@
or false (thus violating bivalence), and more specifically, as obeying intuitionist logic %%@
or a close cousin of it. This idea goes back to his early work, e.g. the closing paragraphs %%@
of (1959), but remained central to his thought---as we will see in Sections \ref{curious} %%@
and \ref{Dummcon}.

\subsection{Work for another day}\label{whatelse}
I admit that my choice of these three topics is biassed, in that I have written on them %%@
before. Indeed, I cut my philosophical teeth on the first of them. On the other hand, I %%@
submit that these topics are not a `stretch'. Despite different prevailing concerns in %%@
philosophy of language and logic, and in philosophy of physics---and despite Dummett's %%@
anti-realism, and my own aspiration  to be a naturalistic realist---all three topics %%@
involve  connections close and substantial enough to be worth stating.

Furthermore, I submit that one could well `play the same game' with other topics Dummett  %%@
addressed. As regards time, there are two obvious topics. One is Dummett's rejecting the %%@
idea that time is composed of instants, in the way that the classical continuum is composed %%@
of real numbers, on the grounds that it admits as conceptual possibilities---a philosopher %%@
of physics might say: kinematic possibilities---motions of a body that are so discontinuous %%@
as to be surely conceptually impossible (2000, 500-505). He goes on to discuss %%@
alternatives. For example:  a moment of time might be modelled by the set of rational %%@
numbers in some open interval of real numbers, smaller than any time-resolution we shall %%@
ever devise; and the value of another quantity such as position will be, not a real number, %%@
but the set of rationals in some open real interval, smaller than any measurement %%@
resolution we shall ever devise.

In response: I applaud the investigation of these alternative models, but cannot pursue it %%@
here. Let it suffice to commend some subsequent discussion (Meyer (2005), Dummett (2005), %%@
Butterfield (2006, Sections 3, 4)), together with other work on the sort of discontinuous %%@
kinematic---and even dynamic---possibilities which Dummett rejects (e.g. Norton 1999, 2008; %%@
Perez Laraudogoitia 2009).

A second obvious topic is causal loops and time travel. Again, I will not here pursue this %%@
alluring idea; but will just make two points, by way of  encouraging further work. First: %%@
the springboard in Dummett's writings is his suggestion  that backwards causation is %%@
coherent, provided that current intentions about whether to perform the action that is a %%@
putative sufficient condition of a past event, can be as good evidence about whether the %%@
past event occurred as are that event's current traces. (For a more precise statement, cf. %%@
his 1964, especially 349-350; cf. also 1954, 327-332, and 1986, 359-362.) I believe that %%@
Dummett's suggestion coheres with the basic idea in most defences of the possibility of %%@
time travel; and that this idea should be uncontroversial. This idea is that time travel %%@
simply imposes a stringent consistency condition on states and their time-evolution, viz. %%@
that the initial state (`a youthful Tim, holding a rifle, disembarking from his %%@
time-machine in his grandfather's home town') must evolve back to itself (and so via a %%@
state like: `a youthful Tim, holding a rifle, embarking on his time machine, intent on %%@
killing his grandfather, the profiteering munitions magnate'). (The example of Tim is from %%@
Lewis (1976, 75-80).)

Second: we note that backwards causation and time travel is a topic, not just in current %%@
philosophy of physics, but also in physics, admittedly in its most speculative reaches! %%@
Good recent work on backwards causation (including as a route to solving the measurement %%@
problem of quantum theory) includes Berkovitz (2008), Kastner (2008) and Price (2008); cf. %%@
also the brief discussion in the appendix to Dummett's 1986 (370).  For a philosopher's %%@
introduction to time travel in physics, cf. Earman et al. (2009),  Smeenk and Wuthrich %%@
(2011).

Besides, time is of course not the only area in which Dummett's writings bear on issues in %%@
physics. One obvious area is `quantum logic', i.e. the proposal that logic should be %%@
revised to incorporate the non-distributive structures in quantum theory (rebutted by %%@
Dummett (1976); relevant recent work includes Bacciagaluppi (1993, 2007), Stairs (2006)). %%@
Another is scientific realism and the perceptual basis of empirical knowledge, addressed by %%@
Dummett in his (1979).

\subsection{Time, modality and semantics}\label{tms}
In this Subsection, I will state what I take to be the main philosophical debate about %%@
time, and mention related issues about modality and the semantics of temporal language. Of %%@
course, I will not try to settle the debate or related issues. But they are worth stating. %%@
For the debate and issues will form the backdrop to all three of my connections.

The main philosophical debate about time is the debate about whether or not `temporal %%@
becoming', the `movement of the now', is real.  Jargon varies. Some authors say the debate %%@
is about whether there are `tensed facts' (`tenserism') or not (`detenserism'); some adopt %%@
a notation of McTaggart's (1908), saying the debate is between the `A' vs. `B' %%@
(`block-universe') views of time. I shall adopt the first jargon, tenserism vs. %%@
detenserism.\footnote{There are yet other jargons, e.g. `eternalism' for `detenserism'. %%@
Dummett himself (2004, Chapter 5) notes that there are four possible positions about the %%@
reality of the past and the future---that neither is real, or one but not the other, or %%@
{\em vice versa}, or both are---and suggests labelling them respectively as Model (1), (2), %%@
(3) and (4). So detenserism would be labelled `model (4)'.}

Even the statement of the debate is contentious, some saying that phrases like `temporal %%@
becoming' and `the moving now', are irredeemably vague or metaphorical. But I think the %%@
problem is at worst a matter of the relevant words---`real' and its ilk, like %%@
`objective'---being ambiguous: rather than their being irredeemably vague or metaphorical.

 Thus here is one possible meaning for detenserism; (it is the meaning I will concentrate %%@
on).  Past and future things, events and states
of affairs (or however one conceives the material contents of
spacetime) are just as real as present ones. Abraham Lincoln is just
as real as Bill Clinton, just as Venus is just as real as Earth:
Lincoln is merely `temporally far away from us', just as Venus is
spatially far away.  Similarly for a young child's first grandchild, supposing the child %%@
will have one. And this caveat simply reflects the fact that it is hard
to know about the future (even harder, perhaps, than it is to know
about the past)---not that the future, or its material contents, is of
some different ontological status than the present or past. On the other hand, consider a %%@
contrary tenser doctrine, often called `presentism': that only present things etc. are real %%@
i.e. past and future things etc. are unreal.\footnote{Here, I say
`things etc.' for simplicity: for the main idea of
presentism, it does not matter how you conceive the material
contents of spacetime---though of course in more precise versions, it can
matter. Presentists include, for example, Prior (1970) and Markosian (2004); and we will %%@
see in Section \ref{julian} that Barbour is a kindred spirit.}

So for there to be a clear dispute between detenserism and presentism---or between %%@
detenserism and other rivals, like the idea that present and past things etc., but not %%@
future ones, are real---we need to avoid ambiguities in `is real', `exists' and similar
words. For example,, detenserism should not be just an insistence that we
 use `is real' as short for `has existed or presently
exists or will exist'. And presentism should not be just an insistence on using `exists', %%@
`is real' etc. for
`presently exists'. Rather, we should take it that some
distinction between real and unreal, in intension though not of
course in extension, is common ground to the parties to the
debate. Or at least: it is common ground, as applied to
material things, events etc.; we here set aside mathematical and
other abstract objects.  Then detenserism says, with `real' (or
`exists' etc.) as applied to material things etc.: all past, present and
future things etc. are real.  And presentism says, with the same
sense of `real' (or `exists' etc.): only present things etc. are real.\footnote{I say `with %%@
the same sense of `real'', for
simplicity: it secures a direct contradiction between detenserism
and presentism. But of course different authors can and do make
different distinctions between real and unreal; with the result
that---even if their distinctions are precise---the contradiction
between one man's detenserism and another's presentism can be
much less obvious. Indeed, their choice of distinctions might, at
a pinch, make their positions compatible. For an argument aiming to secure that `real',  %%@
`exists' etc. have univocal meanings, and thus that our debate is genuine, cf. Sider (1981, %%@
xix-xxiv, 16-17). On the other hand: for the view that the debate conflates distinctions  %%@
that in fact cut across one another---promising some compatibilities---cf. Tooley (1997).  %%@
Here we return to the point I mentioned when introducing the meaning of detenserism I will %%@
concentrate on: viz. that there are other meanings. Some (e.g. Maudlin 2007, 126-142) %%@
defend temporal becoming as an objective directedness of the time dimension in a `block %%@
universe' of {\em my} detenser's kind. But Price has given a masterly rebuttal of this kind %%@
of temporal becoming, among others: a rebuttal that combines metaphysics and philosophy of %%@
physics (2011, especially Section 3, 281-302).}

This debate obviously connects in various ways with those about
modality. The principal connection is via using modality to gloss the
real/unreal distinction. Thus `unreal' is often glossed as `merely
possible'. Tensers (i.e. opponents of detenserism) typically say that
the future, and maybe the past, is not actual, but merely possible.
And similarly presentists say (in terms of things, for simplicity):
Abraham Lincoln and Sherlock Holmes are on a par; so are the young child's first
grandchild (supposing there is one---it is hard to know), and Darth Vader (supposed %%@
fictional, as intended!).

This connection with modality means that in recent decades the debate has been
invigorated by developments in modal metaphysics. In
particular, Lewis' bold advocacy of the equal reality of all
possible worlds (1973, Chapter 4.1; 1986) gave a clear modal analogue of
detenserism; and similarly made the contrasting actualist view an
analogue of presentism. Not that these analogies made everything
cut and dried. In particular, as Lewis himself emphasised:\\
\indent (i) one should not just identify `being real' with `being concrete',
since the concrete-vs.-abstract distinction is itself in bad shape
(1986, Section 1.7);\\
\indent (ii) one cannot expect the debates about the identity of items, through time and %%@
across possible worlds, to be strictly parallel---not least because here various
proposed distinctions between things, events, states of affairs etc.
come to the fore. In particular, the {\em pros} and {\em cons} of the doctrine that objects %%@
persist over time by consisting of temporal parts (`stages'), each confined to its time, %%@
may not run parallel to the {\em pros} and {\em cons} of the analogous doctrine that %%@
objects exist in different worlds by consisting of different objects, each confined to its %%@
world (Lewis 1986, Chapter 4.1-4.3).

On the other hand, we
should not assimilate this debate to one in {\em semantics}. Detenserism, presentism and %%@
their ilk are not just rival proposals for the semantics of temporal language. There is a %%@
temptation to see them like this; (indeed, I think the literature of the 1950s to 1970s was %%@
wont to do so; cf. Butterfield (1984)). Thus detenserism seems to go with a simple bivalent %%@
semantics which,
prescinding entirely from all the complexities of natural language,
uses either: \\
\indent (i) a single domain of quantification containing all
objects that ever exist; or \\
\indent (ii) a linear order of domains, each
containing the objects that exist at a single time, so that the
quantifier represents present-tensed `exists'; (here `object' covers
things etc.) \\
With either (i) or (ii), `now' and other temporal indexicals get a
straightfoward time-dependent reference. (For example: If times are
treated as objects in the domain, then `now' can be assigned a time as
reference.)  Correspondingly, tenserism and presentism seem to go with
more complex semantic proposals: say with using three truth-values, or
a branching future; or both of these.

But we should beware of the gap between semantics and metaphysics:
each discipline is, and should be, beholden to considerations,
substantive and methodological, that the other ignores. In the present
context, not only might linguists have reasons for or against these
semantic proposals, which ride free of metaphysics. Also, the
proposals do not straightforwardly express the metaphysical positions,
just because formal semantics is not concerned with what is `real'.

Thus the use of a single big domain of quantification, as on the first
proposal, is not implied by all its members being real; so the
detenser may well endorse one of the more complex semantics. And the
tenser will note that even these proposals do not capture her metaphysical
thesis about reality.  In particular, any such semantics requires
`now' and other temporal indexicals to be treated just as they were in
the simple bivalent semantics.  It is part and parcel of doing
semantics---whether with two truth-values or more, whether with
branching or not---that such indexicals get a straightfoward
time-dependent reference.  So the `movement of the now', which for the
tenser and presentist is the crucial fact about time, is represented
only by the semantics' use of a family of interpretations, related to
each other by `sliding along' the reference assigned to `now'
etc.---exactly as in the simple semantics apparently favoured by the
detenser!\footnote{I believe this point is not affected by the
complexities of allowing for relativization of truth-value to
circumstances of assessment, as well as circumstances of utterance;
but I cannot argue the point here.}

So much by way of introducing themes about time, modality and semantics. It will be clear %%@
how each of my three connections relates, not just to the main debate, but to some of the %%@
specific issues mentioned. In brief: Section \ref{seepres}'s connection relates to %%@
persistence over time, and semantics; Section \ref{split}'s connection relates to how we %%@
understand indexicals; Section \ref{julian}'s connection relates to modality---namely with %%@
a bold idea for naturalizing it along the lines of the Everett interpretation of quantum %%@
theory!

\section{Seeing the present}\label{seepres}
\subsection{Dummett's remarks}\label{Dummremks}
My first connection is based on some remarks in the final Section of Chapter 11 %%@
(`Thoughts') of Dummett's {\em Frege: Philosophy of Language} (1973). The Section, entitled %%@
`Token-reflexive expressions' is about whether Frege's doctrine that thoughts have an %%@
absolute truth-value has to be modified to allow for token-reflexive expressions. It is %%@
long  (382-400), and moves seamlessly from the topic of variable truth-value (especially, %%@
from 385 onwards, with respect to time), to existence, and then to observation, and then to %%@
the semantic analysis of temporal language. But I will focus on his remarks about %%@
observation, on page 388.

Broadly speaking, I will endorse Dummett's remarks---and report how long ago, I was %%@
inspired by them to make some philosophical hay of my own. But I will also note a %%@
disagreement with Dummett's use of the remarks (later in the Section) to argue that in %%@
semantics temporal indicators should be analysed as sentence operators (rather than as %%@
terms standing for times).

Dummett writes: `What we think of as properties of material objects are, typically, things %%@
that can be predicated of them at a given time, and may be false of them at another time. %%@
The reason is quite obvious. The basic predicates of our language, those which we first %%@
learn to employ, are ones whose application can be determined by observation ... [and] %%@
...an observation can determine only how [an] object is at some one time' (388). There is a %%@
disanalogy with space here: observation is not thus restricted spatially. Admittedly, one %%@
may not be able to the whole spatial extent of a very large object at once, but `most %%@
observational predicates apply to an object as a whole considered as it is at a particular %%@
time' (388).

Then in a footnote Dummett adds that `most objects which we observe are  close to us, %%@
relative to the speed of light and to the rate at which we make observations, so that in %%@
practice ... we take observation as revealing the state of the object at the time of %%@
observation ... [besides] ... the primary method of determining the application of an %%@
observational predicate can often be employed over a wide range of distances at which the %%@
object may be placed. Thus for practical purposes, we determine how an object is at a given %%@
time by observing it at that time' (388).

These remarks essentially provide two asymmetries between time and space; (some remarks %%@
which I have omitted give details about this contrast with space). The first asymmetry %%@
(from the main text) is that in order to ascribe most observational predicates we need to %%@
observe the whole object; but as the predicate applies to the object `considered as it is %%@
at a particular time', we do not need to observe the object's entire life-history. The %%@
second (from the footnote) is that usually we can ignore the time-lags involved in %%@
observing distant objects. That is, we can take observation  to inform us of objects' %%@
properties and relations at the time of observation even if they are not at the place of %%@
observation.

\subsection{The asymmetries endorsed---and exploited}\label{asymend}
I endorse both these asymmetries---with some clarifications, and for the same reasons, that %%@
I gave long ago. To avoid repeating those discussions, let me just summarise as follows. As %%@
to the first, my main reason lies in the fact that whatever our attitude to temporal parts %%@
might be, we all accept that objects have {\em spatial} parts which are genuine objects: %%@
people have arms, chairs have legs etc. Thus observational predicates tend to apply to %%@
whole objects---not so much because most objects are small enough, or transparent enough, %%@
to be observed in their entirety---but because, when they are {\em not} entirely %%@
observable, we take a spatial part of the given object, to be the object to which the %%@
predicate really applies. For details, cf. Butterfield (1985, 41-42). Incidentally: Dummett %%@
rejects temporal parts, but in recent work said that the detenser should too (2003, 51-52; %%@
2004, 86-88).

As to the second, my main reason is that, indeed, for the senses of sight, hearing and %%@
touch, most of the objects we observe rarely change their observable properties during the %%@
time-lag involved in the process of observation. (Smell is an exception: we can smell burnt %%@
toast long after the toast has stopped burning. So perhaps is taste.) A similar point %%@
applies to oral communication. We can usually ignore the time-lag in speech, i.e. take the %%@
speaker to (purport to!) believe what he said, at the time the hearer receives and %%@
understands the message, and not merely at the earlier time of utterance. Cf. Butterfield %%@
(1984a, Sections 2,3).

For this second asymmetry, it is also worth adding some details. For discussions often %%@
emphasize only that observation takes very little time, and in particular cite the amazing %%@
rapidity of light; and neglect the equally important issue of how long the observed object %%@
typically keeps the property in question. (For brevity, I shall set aside the corresponding  %%@
points about communication.)

Thus there are two temporal  factors to be considered:\\
\indent (i) the typical time it takes to make an observation, i.e. the time it takes for a %%@
causal chain to leave the object, reach us, pass through our sensory system and finally %%@
yield an observational judgment; and \\
\indent (ii) the typical time-scale on which the observed object keeps its observable %%@
property, i.e. the typical time-interval between  changes in the property.\\
Provided (i) is smaller than (ii), we can (typically!) make a present-tensed observational %%@
judgment: such as `there is a blackbird on the tree in the garden', rather than `there was %%@
(or: was about $N$ seconds ago) a blackbird on the tree in the garden'. In this blackbird %%@
example\footnote{Taken from Dummett (2005, 680): who uses it for a very different %%@
purpose!}, the time (i) is so small---since, in particular, light is so fast---that, %%@
although birds often do not stay long on a tree, we can be confident that (i) is smaller %%@
than (ii), and thus that the present-tensed judgment is true.

So much by way of endorsing Dummett's second asymmetry: in short, that usually we observe %%@
(and in speech, communicate with) the {\em present}, though spatially distant, state of %%@
affairs. But I also believe that this asymmetry leads to convincing explanations of three %%@
other time-space asymmetries that might be, and have been, taken to support the {\em %%@
tenser}. Namely:\\
\indent (i): We more readily take as real the presently existing objects, wherever they %%@
are, than the objects that are at some time located here (e.g. Putnam 1967, Dummett 2003, %%@
34; 2004, 52).\\
\indent (ii): We are more apt to give sentences time-variable truth-values than %%@
space-variable ones (e.g. Dummett 1973, 386, 390). \\
\indent (iii): We think of ourselves as sharing a common, albeit ever-changing, {\em now}, %%@
while we each have a different {\em here} (e.g. Gale 1964, 105).

These explanations are spelt out in Butterfield (1984a, Sections 4-6); and I will not %%@
repeat them. Suffice it to say that they are `naturalistic' appeals to uncontroversial %%@
physics and psychology, and so will be welcomed by the detenser as `explaining away' these %%@
asymmetries. They have also been further developed: Callender finds the third asymmetry the %%@
most compelling (2008, Section 3), and goes on to add many empirical details to my %%@
explanation (2008, Sections 4-6). On the other hand, since the science in these %%@
explanations is uncontroversial, I submit that the tenser should also accept them---and so %%@
has a responsibility  to clarify whether she thinks any of these three asymmetries has a %%@
further content, or significance, which is not captured by these naturalistic explanations %%@
and which expresses part of her tenserism. So far as I know, this responsibility has not %%@
yet been discharged!

Finally, let me clarify that I do not especially intend to press Dummett on this last %%@
point. Despite the citations just given, he did not, so far as I know, urge any of %%@
(i)-(iii) as a straightforward argument for a tenser position. On the other hand, he did %%@
argue (1973, 389f.) that the two asymmetries (from his 388) which I have endorsed,  perhaps %%@
together with related considerations, have consequences for semantics. Namely: they make it %%@
correct, or at least more natural, to analyse temporal indicators as sentence operators, %%@
rather than as terms standing for times. For a detailed critique of this argument, cf. %%@
Butterfield (1984b).\footnote{A `halfway house' semantics, in which  temporal indicators %%@
that qualify singular terms are analysed as predicate modifiers, is developed by %%@
Butterfield and Stirling (1987, Section 4).  For trans-temporal relations as a problem for %%@
the presentist, cf. Sider (2001, 25-28).}

\section{The essential indexical---for branches}\label{split}
I  take my cue from Dummett's discussion of McTaggart's argument for the unreality of time %%@
(Section \ref{McT}). This leads to the indispensability of indexicals to express the %%@
uncertainty before a `splitting' of the universe into `branches', according to the Everett %%@
interpretation of quantum theory (Section \ref{ustdsplit}).

\subsection{McTaggart's argument}\label{McT}
McTaggart's (1908; 1927, Chapter 33) argument forms a cross-roads where several aspects of %%@
the tenser-detenser debate meet, such as: the analogies and disanalogies between time and %%@
space, the relation between time and change, and  the logical behaviour of temporal %%@
indexical expressions, especially `is past', `is present' and `is future'.

Broadly speaking, the argument has two parts. In the first part, McTaggart argues that (i) %%@
time involves change, and (ii) change requires tensed facts (in his jargon: A-series %%@
facts), i.e. the objectivity of temporal becoming. McTaggart's reason for (ii) is that it %%@
is necessary, if change is to be distinguished from spatial variation in properties. In the %%@
second part, McTaggart argues  that tensed facts involve a contradiction. This is  a %%@
regress argument, in which he envisages iterating the temporal indexicals `is past', `is %%@
present' and `is future'.

Dummett's  (1960) defence proceeds as follows. He emphasizes  that there are analogous %%@
regress arguments using spatial indexicals like `is here' and `is there' (or `nearby' and %%@
`far'), or personal indexicals like `I' and `you'; and that since McTaggart `does not ... %%@
display the slightest inclination to apply his argument in this way to space or to %%@
personality' (353), we should focus on the first part of the argument. Dummett endorses %%@
this first part (354-5); and in recent work, he apparently again concurred (2003, 51; 2004, %%@
87-88). He takes its conclusion to be that there cannot be an indexical-free complete %%@
description  of a temporal reality (while there can be such a description of spatial %%@
reality). Here is one of his formulations: `a description of events as taking place {\em in %%@
time} is impossible unless temporally token-reflexive expressions [i.e. indexicals] enter %%@
into it, that is, unless the description is given by someone who is himself in that time' %%@
(353).

Dummett then raises the question how to reconcile this conclusion with the second part of %%@
the argument, and its avowed conclusion that time is unreal. After all, as Dummett says: %%@
the first part of the argument seems to demonstrate `the reality of time in a very strong %%@
sense, since it shows that time cannot be explained away or reduced to anything else' %%@
(356). He suggests a reconciliation. Namely: he thinks that McTaggart is assuming that %%@
anything real must have a complete---that is: observer-independent, or indexical-free--- %%@
description. He ends by raising the worry that McTaggart's conclusion that time is unreal %%@
is self-refuting: for even if the world is really atemporal, our apprehension of it surely %%@
changes. This worry prompts Dummett, in his last paragraph, to toy with applying {\em modus %%@
tollens}---directing it at the assumption he has just attributed to McTaggart. That is: he %%@
toys with the idea of denying that anything real must have an indexical-free description.

I of course cannot address all the themes Dummett raises; let alone other possible %%@
interpretations of McTaggart.\footnote{By my lights, the main rival is Mellor's  %%@
diametrically opposite interpretation (1981, Chapter 6, especially 92f). He rejects the %%@
first part of the argument, and endorses the second as showing temporal becoming (in his %%@
jargon: tensed facts) to be contradictory.} In this Section, I will only develop the idea %%@
of the `essential indexical'; (and Section \ref{julian} will pick up on Dummett's closing %%@
discussion of whether time being unreal is self-refuting).

As we have just seen, Dummett here articulates this idea as specific to time, and as the %%@
conclusion of the argument's first part. But nowadays most philosophers take a more low-key %%@
view. They see the `essential indexical' as applying equally to space and to personality; %%@
and as implying, {\em not} that there are `perspectival facts' in some sense (the temporal  %%@
variety being tensed facts), but only that indexicals are indispensible for conveying the %%@
contents of our thoughts and sentences.\footnote{This view was argued for, wittily and %%@
persuasively,  by Perry (1979) and Lewis (1979); and since then, it has been often endorsed %%@
and developed. Examples in the philosophy of time are: Mellor (1981, Chapter 5, especially %%@
78f), Butterfield (1984c, 77-85) and Sider (2001, 18-21). The view  also has precursors, %%@
whom Perry cites, e.g. Castaneda. Butterfield (1986) gives a definition of content (for %%@
utterances and for propositional attitudes) that strikes a compromise between Perry and %%@
Lewis, who advocate contents that are psychologically narrow and have a relativized %%@
truth-value, and authors like Stalnaker and Evans who, in a `neo-Fregean' way, advocate %%@
contents with absolute truth-values.}

 Broadly speaking, I endorse this low-key attitude to the idea of the essential indexical. %%@
But I want to report how the Everett interpretation of quantum theory yields a novel %%@
application of the idea. For it is an application with three features that might appeal to %%@
Dummett. First: it meshes somewhat with Dummett (1960)'s taking essential indexicality to %%@
be about  time rather than space. For some of the considerations about the future being %%@
open etc. that prompt Dummett's view, are endorsed by the Everettian branching: roughly %%@
speaking, the open future is understood as the effective, but not fundamental, %%@
indeterminism associated with the `collapse of the wave packet'. Second: it is %%@
metaphysically revisionary, albeit in a very different way than Dummett's anti-realism. %%@
Third: to understand it, one needs considerations drawn from the philosophy of %%@
language---which the philosophy of physics literature, to its credit, has already deployed.

\subsection{How should we understand branching in the Everett %%@
interpretation?}\label{ustdsplit}
For reasons of space, I will assume familiarity with the basic ideas of:\\
\indent (i): the quantum measurement problem: viz. the apparent conflict between (a) the %%@
continuous and deterministic evolution of quantum states by the Schroedinger equation, %%@
which tends to create superposition states without definite values for physical quantities %%@
such as position, momentum etc., and (b) macroscopic objects' apparently definite values %%@
for position etc.; \\
\indent (ii): the Everett interpretation (Everett 1957): viz. the state of the universe as %%@
a whole is a superposition corresponding to many different definite macroscopic realms %%@
(`macrorealms' or `worlds' or `branches'), which differ among themselves about quantities' %%@
values, e.g. about the positions of the various macroscopic objects; and we should explain %%@
our experience of a single 
definite macrorealm, by postulating that the various macrorealms are all actual---we just %%@
happen to be in one rather than any of the others.

Agreed, these ideas call out for philosophical clarification. My own attempts were (1995, %%@
1996, 2002a); the state of the art is represented by Saunders et al. (2010), and Wallace %%@
(2012), both of which are outstanding. But here I will focus just on the issue to which the %%@
idea of the essential indexical applies. This is the problem of probability: (more %%@
precisely, the `qualitative problem of probability'---to distinguish it from a quantitative %%@
problem, which is whether the Everettian can justify the values of the orthodox quantum %%@
probabilities).

The problem is that probability seems to make no sense, if all possible outcomes of a %%@
putatively probabilistic process in fact occur---as the Everettian says they do. For %%@
according to the Everettian, the quantum state always evolves deterministically, even %%@
during quantum measurements and the other processes such as radioactive decay, that are %%@
traditionally taken
as indeterministic `collapses' of the quantum state into just one of various possible %%@
outcome states. Thus the Everettian says that during such a process, the quantum state %%@
evolves to include a term (i.e. a summand in a sum) for each possible outcome, and that the %%@
universe splits into many branches, in each of which one of the outcomes occurs.

I think that all Everettians (both nowadays and yesteryear) should agree that the answer to %%@
this problem must lie in invoking {\em subjective uncertainty}. The basic idea will be an %%@
analogy with how probability is taken as subjective uncertainty, for a deterministic %%@
process of the familiar classical kind. For such a process, a unique future sequence of %%@
states is determined by the present state (together with the process' deterministic law). %%@
But the agent or observer does not know this sequence in advance, either because she does %%@
not know the present state in full detail or because she finds it too hard to calculate the %%@
future sequence from the present state.

Similarly, says the Everettian: probability can be taken as subjective uncertainty, for a %%@
deterministic process of the {\em unfamiliar} Everettian kind. For such a process, a unique %%@
future sequence of `global' states is again determined by the present quantum state %%@
(together with the Schroedinger equation). And here, unlike the classical case, one can %%@
assume the agent or observer does know the present state, and how to calculate from it the %%@
future sequence. But the agent or observer is nevertheless uncertain since, thanks to the %%@
impending `branching' or `splitting', she will not experience any such future `global' %%@
state, i.e. she will not experience the outcomes corresponding to all its terms. At each %%@
future time, she will only experience one outcome---and is thus uncertain about which.  %%@
Thus this kind of uncertainty, compatible with full knowledge of the global state and the %%@
laws, is rather like the self-locating uncertainty discussed by philosophers under the %%@
heading `the essential indexical' (cf. again Perry 1979, Lewis 1979).

But the phrase `rather like' papers over a debate about the exact nature of this %%@
uncertainty. I will only report the recent views of Wallace: views which I favour, and %%@
which have the merit, for connecting to Dummett's views, of invoking considerations in the %%@
philosophy of language.  So what follows is a glimpse of Wallace (2005; 2006; 2012, Chapter %%@
7). (Note: Wallace emphasizes that his views develop earlier work by Saunders. For %%@
discussion, including contrary views, I also recommend e.g. Greaves (2004); and the debate %%@
between Saunders and Wallace and Tappenden---for references, cf. the latest round, %%@
Tappenden (2011).)

It will help to focus, as the literature often does, on as simple a case of Everettian %%@
branching as possible. For example, consider a quantum measurement with just two possible %%@
outcomes, say `up' and `down' for a measurement of spin (which is a two-valued quantum %%@
quantity). We also want to set aside the sort of uncertainty about the future that arises %%@
even under classical determinism: namely, when an agent or observer does not know the %%@
present state in all its details and-or lacks the skill to calculate the future state from %%@
all those details.  So to focus better on what is distinctive of the Everettian case, we %%@
envisage an observer of the quantum measurement, Anna, who knows all the relevant %%@
details---the prior state of the quantum system being measured, the details of the %%@
apparatus etc.---and suffers no limitations about calculating. In particular, Anna can %%@
calculate the two outcomes' orthodox quantum
probabilities. (`Probabilities', as we call them! Of course, the Everettian's right to call %%@
them that is what is at issue.) We may as well take the outcomes to be equi-probable: each %%@
with probability one-half.

So: what should Anna's attitude be before the measurement? There are two rival lines of %%@
thought: intuition pulls in two directions. I shall follow Saunders and Wallace in %%@
endorsing the first line. But of course, my aim is not to contribute to the debate; to do %%@
that in so small a space, for a debate so vigorous,  would be a tall order! I aim only to %%@
summarize Wallace's views, so as to exhibit: (i) the appeal of the `essential indexical' %%@
and (ii) the role of considerations from the philosophy  of language.

So here is the first line of thought. Anna  should feel uncertain of the outcome, despite %%@
her knowledge of all the physical details, and even if she is a convinced Everettian. For %%@
she will not observe both outcomes. Rather: according to the Everettian (and Anna herself, %%@
if she is convinced) Anna, along with other emergent macro-objects like the apparatus and %%@
its pointer, will split in the course of the measurement, some of her successors seeing %%@
`up', and some seeing `down'---and of course, some others seeing no outcome because the %%@
measurement goes wrong, or they faint during the measurement, or they slip on a banana-skin %%@
and get concussed, or ... . But we can leave these unfortunates aside: they do not affect %%@
the ensuing argument.\footnote{Of course, in a well-designed experiment, in a well-run lab %%@
(without banana-skins!), these unfortunates will have low quantum probability, and would %%@
therefore also be discounted in many other discussions, e.g. of the confirmation of quantum %%@
theory.}  So we can take it that any successor sees `up', or sees `down'.  And of course, %%@
no successor sees both outcomes---although both occur in the Everettian multiverse. Hence %%@
the uncertainty before the measurement.

Hence also the idea that the proposition about which Anna is uncertain is indexical, or %%@
self-locating, in the sense of the phenomenon of the essential indexical: though of course, %%@
the `dimension' of indexicality is---not one of that familiar trio, space or time or %%@
personality, but---what one might call `branchness'.  However, it is natural to express the %%@
indexicality with `I' and similar words like `my successor', since our discussion is %%@
concerned with a person's uncertainty, and we naturally envisage that the time and place of %%@
the measurement can be robustly identified between the different Everettian branches and %%@
can be known by the person. Thus Anna, sitting with eyes closed at 11.59 at the Cavendish %%@
Lab., Cambridge, waiting to open them at noon in order to see the outcome, will find it %%@
natural to say: `I wonder whether  I [or: my successor], at noon in Cambridge, will see %%@
`up''. (But cf. Butterfield (1995, 141-142): which (i) warns that, though  this use of `I' %%@
is natural, it does not imply metaphysical doctrines about personal identity, and (ii) %%@
points out that we can indexically specify branches by `this branch' etc.)

I endorse the intuitions of the last two paragraphs; (following Saunders and Wallace). But %%@
on the other hand, I admit that it violates some apparently compelling principles about %%@
knowledge, expectation and uncertainty. Thus Greaves (2004) states two such:
\begin{quote}
... whatever she knows she will see, she should expect (with certainty!) to see. So she %%@
should (with certainty) expect to see spin-up, and she should (with certainty) expect to %%@
see spin-down. (Not that she should expect to see both: she should expect to see each.) %%@
(440) ... I [meaning in this context: any rational person] can feel uncertain over $P$ only %%@
if I think there is a fact of the matter regarding $P$ of which I am ignorant (441).
\end{quote}
Greaves goes on to develop her own account of Everettian branching, which has come to be %%@
called `the fission programme'; as against Saunders and Wallace's `subjective uncertainty %%@
programme'.

I will not go into further details about this dispute. Suffice it to make three points; the %%@
second and third will be the positive and important ones. First: Wallace replies to the %%@
apparent appeal of  the principles formulated by Greaves. In particular, he diagnoses an %%@
ambiguity between: a true non-technical interpretation, which is no problem for the %%@
Saunders-Wallace subjective uncertainty view; and a contentious technical one; (cf. %%@
especially his (2006, 667f; 2012, Chapter 7.6.)

Second: Wallace (especially 2005; 2012, Chapter 7.3-7.5) spells out two rival proposals  %%@
for the semantics of a language spoken by people in a universe subject to Everettian %%@
branching; and he considers the case where they know the universe branches, and the case %%@
where they do not---and the case where they discover that it does! (Wallace says `model' %%@
not `proposal': but I prefer the latter, since `model' has many other uses, especially in %%@
semantics!) I will only need the main contrast between these proposals, not their details. %%@
For that contrast will be enough to show that the proposals underpin, respectively, the %%@
rival intuitions about Anna's state of mind, her certainties  and uncertainties, and about %%@
what we should say about them. My third point will then be about how to choose between the %%@
proposals.

Both proposals assign to sentences truth-conditions (not conditions of assertibility or %%@
justification---apologies to Professor Dummett!) in terms of `possible worlds', i.e. %%@
roughly, total possible courses of history.\footnote{But the proposals may be closer to %%@
Dummett's concerns than meets the eye, largely because they are focussed on future %%@
uncertainty. Thus Wallace surmises (personal communication) that they could be restated, %%@
{\em mutatis mutandis}, in terms of conditions of assertibility or justification.} The %%@
contrast comes in how each conceives of a possible world. For the first proposal, a %%@
possible world is an Everettian branch. For the second proposal, a possible world is a %%@
trajectory through the quantum mechanical state-space, i.e. a specification for each time %%@
of the quantum state of all the systems concerned---in principle, the quantum state of the %%@
whole universe. So on the second proposal, a possible world is an entire branching %%@
structure, with all its branches.

This contrast is enough to show how the proposals will differ radically on propositions %%@
about the future---and so will line up, respectively, with the two rival intuitions about %%@
Anna. Thus consider the first proposal. It will function like a branching semantics of the  %%@
sort often advocated by semanticists (especially  tense-logicians) and metaphysicians to %%@
model the idea of `fixed past, open future'. The simplest version of such a semantics will %%@
say that a proposition, evaluated at a time $t$ (where `time' denotes a node in the tree, %%@
rather than a `rank', i.e. set of nodes at the same `height'), which is about the future, %%@
is true just in case what it says holds good on every branch through $t$. For example: let %%@
the proposition be, in the usual tense-logical notation, $F_n p$, `It will be the case in %%@
$n$ units of time that $p$'. Then the proposition $F_n p$ is true at the node $t$ provided %%@
$p$ is true at all nodes that are $n$ units to the future of  $t$.

Let us apply this to the case of Anna's measurement, with $t$ being 11.59---i.e. before the %%@
measurement, in a branch defined informally by the past macroscopic (`coarse-grained') %%@
history of Anna, her apparatus, her laboratory, her environment ... the universe---and with %%@
the measurement being completed at noon. We get the following verdicts:\\
\indent (i): `the spin will be up' and `the spin will be down' are both false at $t$; since %%@
each fails to hold in some of the future branches through $t$;\footnote{It is tempting to %%@
say, not just `fails to hold in some branches', but also `fails to hold in about half of %%@
the branches'. This is legitimate---provided that (i) the Everettian has solved what I %%@
earlier called the quantitative problem of probability, i.e. has justified invoking the %%@
orthodox  quantum probability-measure, and (ii) the orthodox probabilities for the two %%@
outcomes are indeed about one-half.}   \\
\indent (ii): `the spin will be up or the spin will be down (but not both)' is true at $t$.

Agreed, we have obtained these verdicts from what is merely the simplest truth-clause for a %%@
future-branching semantics. There are two points here. First, I have followed Wallace's %%@
(2005) proposal that sentences ascribed truth-values relative to a time, i.e. a node in the %%@
tree; whereas he prefers now (2012, Chapter 7.3-7.5) to ascribe truth-values relative to a %%@
branch. But I shall not pursue this contrast.  

Second, there are future-branching tense-logics that model the idea of a single actual %%@
future, with respect to which future-tense propositions are to be evaluated. Namely, they %%@
single out, relative to every node $t$ of the tree, one future branch through $t$: it %%@
represents the `actual' future in that the truth-clauses for $Fp, F_n p$ refer only to it, %%@
not the other future branches through $t$.

But for present purposes, we should stick to this simplest proposal, for two reasons. %%@
First, for Everettian branching, all the future branches through $t$ are equally real: %%@
recall that we, and our branching semantics, are not trying to model  indeterminism or %%@
stochasticity as ordinarily or classically understood. Second, recall that our overall aim %%@
is to model---to better understand---Anna's certainties and uncertainties, and so her %%@
degrees of confidence, encoded in her subjective probabilities. Thus in order to clarify %%@
the contrast with the case of classical indeterminism, we  assumed that she knew all the %%@
details of the initial quantum state, and suffered no calculational limitations. %%@
Accordingly, since high subjective probability controls assent to propositions ({\em %%@
ceteris paribus}, and allowing for Gricean rules of conversation), it is to the {\em %%@
credit} of this simple proposal that its verdicts for truth and falsity match Anna's %%@
certainties and uncertainties: for example, her willingness at 11.59 to assent to (or even %%@
assert) `the spin will be up or the spin will be down (but not both)', and her then %%@
rejecting both `the spin will be up' and `the spin will be down'.

On the other hand, let us now consider the second proposal. According to this, a possible %%@
world is uniquely specified by an initial quantum state of the systems concerned (in %%@
principle, the universe), and its deterministic (i.e. orthodox unitary) time-evolutes. So a %%@
possible world comprises an entire branching structure. As a result, the semantics in terms %%@
of worlds, thus defined, does {\em not} branch: it is linear.  Although---assuming Anna's %%@
measurement is correctly performed---there will be branches future to $t = 11.59$ with spin %%@
up, and also branches with spin down, all these branches are contained in the same possible %%@
world. Thus the truth-clause for propositions about the future will be `blind' to worlds' %%@
inner structure, and so we get the following verdicts:\\
\indent (i'): `the spin will be up' and `the spin will be down' are both true at $t$; since %%@
each  holds in some  of the future branches through $t$; (again, we might say `about half %%@
the branches'---cf. footnote 11);  \\
\indent (ii'): `the spin will be up or the spin will be down (but not both)' is false at %%@
$t$.

Third, and finally: How to choose between these proposals? This is where, according to %%@
Wallace, philosophy of language---as against semantics or logic---enters the arena. (At %%@
last, I fulfill my promise to connect with this area of Dummett's interests!) More %%@
specifically: principles of interpretation  enter, in particular the principle of charity; %%@
or perhaps better,  the principle of humanity.
For present purposes, we need such principles only in a very rough form, such as the %%@
following.
First, charity: we should so interpret people's words that the beliefs we thereby take them %%@
to express are (by our lights!) mostly true. Or perhaps better, as proposed by the %%@
principle of humanity: their beliefs come out as mostly true (by our lights!), except when %%@
we can explain their error, for example by their not having as much evidence as we do.

Thus Wallace argues that these principles clearly favour the first proposal, especially for %%@
the interpretative situation in which {\em he} finds himself: namely that of a convinced %%@
Everettian whose task is to interpret the speech and behaviour of others, such as Anna, who %%@
conduct quantum measurements. He sees that these people profess uncertainty about %%@
measurement outcomes, i.e. they assent to and reject propositions, very much along the %%@
lines of (i) and (ii), rather than along the lines of (i') and (ii'). Thus charity (or %%@
humanity) dictates that we favour the first proposal over the second.

\section{The reality of the past?}\label{julian}

\subsection{A curious similarity}\label{curious}
My third connection between Dummett's writings and the philosophy of physics is a curious %%@
similarity  between an idea he formulates---roughly, that statements about the past, if %%@
true at all, are true only in virtue of present traces---and the denial of time by Julian %%@
Barbour, the physicist and historian of physics (1999).

To be sure, there are three crucial differences between the two ideas. First: Dummett's %%@
views about the idea---which I will follow him in calling, for short: that the past is %%@
unreal---varied over the years, as Dummett explored the issues (1969; and in more detail in %%@
2003, 2004). On the other hand, Barbour gave a canonical statement  of his denial of time %%@
in his (1999).

Second and more important: as will be clear below, none of Dummett's formulations are %%@
exactly Barbour's doctrine. In summary, the main difference  is that: \\
\indent (i) according to Dummett's idea: all states of affairs about spatiotemporally %%@
localized subject-matters are unreal, except (a) those  that happen to be now known to hold  %%@
(or not to hold), and (b) those whose holding good or not is now effectively decidable; %%@
while \\
\indent (ii) according to Barbour: all states of affairs about spatiotemporally localized %%@
subject-matters are equally real---or what comes to much the same thing: equally %%@
unreal!\footnote{I have stated the ideas in terms of states of affairs. But nothing hangs %%@
on this jargon. I could have spoken of things or events or facts; compare Section %%@
\ref{tms}'s shorthand, `things etc.', for however one conceives the material contents of %%@
spacetime. I have also stated them `ontically, not semantically', as Barbour but not %%@
Dummett would tend to. Dummett would speak of statements e.g. about the past being neither %%@
true nor false, except (a) those  that happen to be now known to be true or false, thanks %%@
to present evidence (traces), and (b) those that are  effectively decidable. But again, I %%@
believe that nothing hangs on this way of stating the ideas.}

Agreed, that summary is indeed obscure---but it will be clear by the end of  Section %%@
\ref{julian2}! And in any case, there is another difference, which is already %%@
comprehensible. It concerns past-future symmetry. Barbour's denial of time is the same for %%@
the future as for the past. But Dummett's idea tends to condemn the past to a more endemic %%@
unreality than the future. For on the one hand, singular observational statements about the %%@
future seem effectively decidable---we naturally envisage making an expedition to the place %%@
and time in question, with instruments, if need be, in hand. On the other hand, for %%@
analogous statements about the past, there is no such procedure: although we can scrutinize %%@
all the present evidence (traces), this gives no guarantee of getting any evidence either %%@
way about the statement in question.

Third: Dummett was clear that he did {\em not} believe in the unreality of the past---nor %%@
did he wish to. He focussed on it just because it seems, worryingly, to be implied by his %%@
advocacy of truth as justifiability; i.e. by what in yesteryear was usually called %%@
Dummett's `anti-realism', which later he called `justificationism': more details in Section %%@
\ref{Dummcon}. On the other hand, Barbour is clear that he {\em does} believe his denial. %%@
And, at least so far as I know---and to the extent that we may `speak with the vulgar' %%@
about whether his views change in a time that he denies!---he has endorsed this statement %%@
since then, e.g. in his (2006, 149-152).

Nevertheless, I submit that the two ideas are similar enough to be worth putting beside %%@
each other---thereby inviting the reader to make a comparison. And fortunately, although %%@
Barbour's view is Everettian in some respects, it will be possible (and clearer) to state %%@
it using only a broad idea from classical physics, especially mechanics. We will need only %%@
the idea of instantaneous states of the system (the universe!) being given by %%@
configurations, such as arrangements in space of various point-particles. But I should %%@
stress here that, to find any {\em reasons} for  Barbour's view, as against just stating %%@
it, one has to turn to quantum physics: more specifically, to Barbour's interpretation of %%@
an approach to quantizing general relativity, called `quantum geometrodynamics'. That is a %%@
complex and controversial subject within physics, and I set it aside completely (cf. %%@
Butterfield (2002, Section 3.2) for a discussion), except to repeat the health warning I %%@
gave in Section \ref{intro}. Namely: Barbour's reasons are more controversial than the %%@
Everett interpretation: indeed, I would say they are idiosyncratic.

I will again first  adumbrate Dummett's discussions (Section \ref{Dummcon}). Then in %%@
Section \ref{julian2}, I state Barbour's view.

\subsection{Dummett's discussions}\label{Dummcon}
In his (1969), Dummett formulated a kind of anti-realism about the past. It is based on his %%@
over-arching theme: that empirical statements that are not effectively decidable violate %%@
bivalence, and should obey intuitionistic logic or a close cousin of it.

Thus the opening point is the fact that statements, or at least most statements, about the %%@
past are undecidable. Though we can of course search for evidence for or against the %%@
statement,  we are not guaranteed to find any evidence, let alone evidence we would %%@
consider conclusive. Besides, we are pretty sure that for countless (`most') statements %%@
about the past, even about straightforward observational matters, we will never have %%@
evidence for or against it. For example, consider: `it rained on the battlefield of %%@
Hastings, eleven days before the battle in 1066'. Or (if the chroniclers recorded the %%@
weather much more assiduously than I imagine they did): `there was once a {\em %%@
Tyrannosaurus rex} where Nelson's Column now is'. This seems to imply that statements about %%@
the past are neither true nor false, except (a) those that happen to be now known to be %%@
true or false, thanks to present evidence (traces, including memories), and (b) those few %%@
(if any) that are  effectively decidable. For, these exceptions aside, there is nothing %%@
{\em now} in virtue of which they can be true.

But as mentioned in Section \ref{curious}, Dummett never welcomed this conclusion. He says %%@
it is `to me and surely to most people ... repugnant: it involved ...
that past events, the memory of and evidence for which had dissipated, were expunged, not %%@
merely from our knowledge, but from reality itself' (2005a, 672; cf. also 2004, 45).

This is not the place to assess Dummett's attempts to avoid the conclusion. Suffice it to %%@
report, as an advertisement for Dummett, a reason he gives why `the justificationist cannot %%@
make it a criterion for the truth of a statement that we possess the means of verifying it' %%@
(2005a, 674). The reason lies in the fact that `truth is what is transmitted from the %%@
premises of a valid argument to its conclusion' (ibid.); and there are countless cases %%@
where we have the means to verify, and even have verified, the premises, but we lack the %%@
means to verify the conclusion. Dummett gives the example of Euler's famous argument %%@
(theorem) that anyone walking across all the bridges of  Koenigsberg must walk across at %%@
least one bridge more than once. Thus: `we can easily conceive of observers stationed at %%@
each bridge, each of whom leaves his post as soon as he sees the walker crossing that %%@
bridge but reports only later without giving the time of crossing; we have then no means of %%@
identifying a bridge he has crossed twice' (2004, p.44; also 2003, 27; 2005a, 674).

From this, Dummett concludes: `[one] must therefore retreat to saying that an empirical %%@
statement is true if it {\em could have been} verified (2005a, 674; cf. also 2004, 45, 92). %%@
He also remarks: `this conclusion ... must come as a relief to anyone attracted to such an %%@
account of meaning and yet troubled about the reality of the past' (2004, 44). Thus for %%@
him, the task becomes one of stating and defending an exact construal of `could have been %%@
verified', and similar phrases, that secures the reality of the past, yet {\em avoids} %%@
collapsing into the opposing realist view (in particular, endorsing bivalence). For that %%@
task and for assessment whether he succeeds, I recommend, in addition to Dummett's %%@
writings, the critiques by Peacocke (2005) and
Moretti (2008).

\subsection{Spontaneity and time capsules}\label{julian2}
I begin with a doctrine I will call `Spontaneity'. For explaining (but not endorsing!) it %%@
will help me state Barbour's view. (It will be yet another sense in which one might `deny %%@
that time is real'.) 

Spontaneity presupposes the idea of a set of many possible
courses of history, where each course of history is a `block
universe' {\em a la} detenserism. But Spontaneity then proposes that unbeknownst to us, the %%@
actual history jumps between disparate instantaneous states.

To explain this, let us suppose we are given, either in metaphysics or in physical
theory, a set of possible courses of history. Setting
aside for a moment the debate between detensers and tensers, we naturally think of
one of these as `real' or `actual' (also: `realized' or `occupied').
And---especially in physics, if not metaphysics---we think of these
possible histories, including the actual one, as continuous in time.
That is, we think of a possible history as a sequence of instantaneous
states of the world (in metaphysics) or of the system (in physics);
and we think of the set of all possible instantaneous states as having
a topology, or some similar `nearness-structure', so that it makes
sense to talk of states being close to each other.  And because, as we
look about us, we seem to see the state of the world changing
continuously, not in discrete jumps, we naturally think that the
possible histories should be not merely sequences of instantaneous
states, but continuous curves in the (topological or similar) space of
such states. So we think of a collection of curves, each curve representing a possible %%@
course of history; and we think of one such curve as real, as actual.

Now I can state Spontaneity more fully. It denies that the possible
histories (including the actual one) need to be continuous in
this sort of sense, and even that `larger' discontinuous changes need be less probable. It %%@
urges that the possible histories, in particular
the actual one, jump about arbitrarily in the space of
instantaneous states.

At first sight, this mind-bending doctrine seems flatly incompatible with
our impression that the state of the world changes continuously.
But it might just be compatible. For the advocate of Spontaneity
will argue that our evidence for that impression---indeed, all
evidence for all empirical knowledge!---consists ultimately in
correlations between experiences, memories and records that are
defined at an instant. Thus: a present observation is not checked
against a previous prediction, but rather against a present
record of what that prediction was; (cf. Bell 1976a,
95; 1981, 136). This predicament, that epistemologically we are `locked in the present', %%@
implies that any jumps of the type that Spontaneity advocates would not be
perceived as such. Immediately after the jump, the new
instantaneous state, at which the actual history has arrived,
contains records fostering the illusion that the state in the
recent past was near (in the topology of the state-space) {\em
it}---and so not near the actual predecessor, which is now a jump
away.

I can now state the
essentials of Barbour's denial of time. In short, it is a
hybrid of Spontaneity and a strong realism about all the possible
instantaneous configurations---a realism analogous to Lewis' well known
realism about all possible worlds (1986).\footnote{Another way to think of it is that %%@
Barbour's view is a hybrid of presentism and a Lewis-like realism about all the possible
instantaneous configurations.} Here, as mentioned at the end of Section \ref{curious}: %%@
`configurations' means, roughly, `state', in a sense appropriate for a mechanical theory, %%@
e.g. the instantaneous arrangements in space of all the point-particles in the system.

Barbour proposes to go further than Spontaneity's denial that the
actual history is continuous. He denies that there is an actual
history (either past or future): there is {\em just} the space of all
possible instantaneous states of the universe. So here `all
possible configurations' does not mean all logically
possible configurations, but rather `all the configurations of our mechanical theory'. The %%@
set of them is called `the configuration space' (Barbour is a Machian; so for him, the %%@
mechanical theory will use relative  configurations: but we can ignore this aspect of his %%@
views.)

And on the other hand, Barbour takes these configurations to be all
equally real, just as Lewis holds the various possible worlds (i.e.
possible courses of history) to be equally real. He of course concedes
that one can mathematically define sets of configurations; and in
particular continuous curves (since the configuration space  will presumably have a
topology), and even curves that obey some laws of motion, e.g. as given by a `least action'  %%@
principle, as in mechanics. But these sets and curves are `just
mathematical': there is no actual physical history faithfully
represented by one of the sets---not even ({\em \`{a} la} Spontaneity) by
a discontinuous set.

That is Barbour's core idea. He obviously needs (as did Spontaneity) to explain away our %%@
impression that there is history, and a
continuous one to boot. More specifically, he needs to argue that we are
epistemologically `locked in the present'; and that the content
of any perception that requires temporal duration (e.g.
motion-perception) is, despite appearances, false.

 Barbour (1999) goes part of
the way to doing that. In particular, as regards the second issue---the
delusiveness of motion-perception---he takes (what we call!)
motion-perception, e.g. of a kingfisher flying over a pond, to involve
the brain containing a whole collection of (what we call!) records of
configurations of the kingfisher and the water. But not just any
collection. Not only are these configurations similar, i.e.  near each
other in some topology or metric on configurations; also, they can naturally be given a %%@
linear order,
so that they correspond to points along a curve in the configuration
space; (1999, 29-30, 264-267). 

So according to Barbour, our impression that there is history arises
from some configurations of the universe (including those we are part
of) having a very special structure: namely, they `contain mutually
consistent records of processes that took place in a past in
accordance with certain laws' (1999, 31). More precisely, they contain
subconfigurations that falsely suggest such a past.  Barbour has a
memorable name for such configurations; he calls them {\em time capsules}.
So in short: a time capsule is any instantaneous configuration that
encodes the appearance of history, for example a history of previous
motion; and Barbour proposes that time capsules explain away our impression that there {\em %%@
is} history.

So much by way of summarizing Barbour's view. As I have hinted, I myself give it no %%@
credence; (my (2002) gives more details). But I commend it to metaphysicians of time as a
vision to contemplate---and to rebut or endorse! And I commend it especially to admirers of %%@
Dummett's `anti-realism', with an invitation to compare it to the anti-realist view of time %%@
that his writings explored. \\ \\

{\em Acknowledgements}: I am very grateful: to Hugh Mellor for encouragement, and %%@
conversations about time, over the years; and to Adam Caulton and David Wallace for %%@
comments on a previous version. This work was supported in part by a grant from the %%@
Templeton World Charity Foundation, which I gratefully acknowledge.

\section{References}
Bacciagaluppi, G. 1993. Critique of Putnam's Quantum Logic. {\em International Journal of %%@
Theoretical Physics} 32: 1835-1846.

Bacciagaluppi, G. 2009. Is Logic Empirical?. In {\em Handbook of  Quantum Logic}, ed. D. %%@
Gabbay, D. Lehmann and K. Engesser,  49-78. Amsterdam: Elsevier.  (also at %%@
http://philsci-archive.pitt.edu/archive/00003380/).

Barbour, J. 1999. {\em The end of time}. London: Weidenfeld and Nicholson.

Barbour, J. 2006. Time and the deep structure  of dynamics. In {\em Time and  history;
 Proceedings of 28th International Wittgenstein Conference}, ed. F. Stadler, M. Stoeltzner, %%@
133-153. Vienna: Ontos Verlag.

Bell, J. 1976. The Measurement Theory of Everett and de Broglie's
Pilot Wave. In {\em Quantum mechanics, determinism, causality and
particles}, ed. M. Flato et al., 11-17. Dordrecht: Reidel.
Reprinted in Bell 1987; page references to reprint.

Bell, J. 1981. Quantum Mechanics for Cosmologists. In {\em Quantum
gravity II}, ed. C. Isham, R.Penrose and D. Sciama, 611-637. Oxford: Clarendon
Press.  Reprinted in Bell 1987; page references to
reprint.

Bell, J. 1987. {\em Speakable and unspeakable in quantum mechanics},
Cambridge: Cambridge University Press;  second edition 2004, with an introduction by Alain %%@
Aspect.

Berkovitz, J. 2008. On predictions in retro-causal interpretations of quantum mechanics. %%@
{\em Studies in the History and Philosophy of Modern Physics} 39: 709-735.

Butterfield, J. 1984. Prior's Conception of Time. {\em Proceedings of the Aristotelian %%@
Society} 84: 193-209.

Butterfield, J. 1984a. Seeing the Present. {\em Mind} 93: 161-176. (Reprinted in: {\em %%@
Questions of time and tense}, ed. R. LePoidevin, Oxford: Oxford University Press, 1998.)

Butterfield, J. 1984b. Dummett on Temporal Operators. {\em The Philosophical Quarterly} 34: %%@
31-43.

Butterfield, J. 1984c. Indexicals and Tense. In {\em Exercises in analysis}, ed. I. %%@
Hacking, 69-87. Cambridge: Cambridge University Press.

Butterfield, J. 1985. Spatial and Temporal Parts. {\em The Philosophical Quarterly} 35:32- %%@
44. (Reprinted in: {\em Identity}, ed. H. Noonan, International Research Library of %%@
Philosophy, Aldershot: Dartmouth Publishing, 1993.)

Butterfield, J. 1986. Content and Context. In {\em Language, mind and logic}, ed. J. %%@
Butterfield, ,91-122. Cambridge: Cambridge University Press.

Butterfield, J. 1995. Worlds, Minds and Quanta. {\em Aristotelian Society Supplementary %%@
Volume} 69: 113-158.

Butterfield, J. 1996. Whither the Minds?.  {\em British Journal for the Philosophy of %%@
Science} 47: 200-221.

Butterfield, J. 2002. The End of Time?  {\em British Journal for the Philosophy of %%@
Science}, 53: 289-330; gr-qc/0103055; PITT-PHIL-SCI00000104

Butterfield, J. 2002a. Some Worlds of Quantum Theory. In {\em Quantum mechanics} %%@
(Scientific Perspectives on Divine Action vol 5), ed. R.Russell, J. Polkinghorne et al.,  %%@
111-140. Vatican Observatory Publications. At: quant-ph/0105052; PITT-PHIL-SCI00000204.

Butterfield, J. 2006. Against {\em pointillisme} in mechanics', {\em British Journal for %%@
Philosophy of Science} 57: 709-754.  Available at: %%@
http://philsci-archive.pitt.edu/archive/00002553/ or
    http://arxiv.org/abs/physics/0512064.

Butterfield, J. and Stirling, C. 1987. Predicate Modifiers in Tense Logic. {\em Logique et %%@
Analyse} 117: 31-50.
	
Callender, C. 2008. The common now.  {\em Philosophical Issues} 18: 339-361. At: %%@
http://philsci-archive.pitt.edu/3656/ and at %%@
http://philosophyfaculty.ucsd.edu/faculty/ccallender/

Dummett, M. 1954. Can an effect precede its cause? {\em Aristotelian Society Proceedings}, %%@
supplementary volume, 28: 27-44. Reprinted in Dummett (1978); page reference to reprint.

Dummett, M. 1960. A defence of McTaggart's proof of the unreality of time.  {\em %%@
Philosophical Review} 69: 497-504. Reprinted in Dummett (1978); page references to reprint.

Dummett, M. 1964. Bringing about the past. {\em Philosophical Review} 73: 338-359. %%@
Reprinted in Dummett (1978); page reference to reprint.

Dummett, M. 1969. The reality of the past. {\em Proceedings of the Aristotelian Society} %%@
69: 239-258. Reprinted in Dummett (1978); page reference to reprint.

Dummett, M. 1973. {\em Frege; the philosophy  of language}. London: Duckworth; second %%@
edition (with the same pagination) 1981.

Dummett, M. 1976. Is logic empirical?. In {\em Contemporary British philosophy}, ed. H. D. %%@
Lewis, London. Reprinted in Dummett (1978).

Dummett, M. 1978. {\em Truth and other enigmas}. London: Duckworth.

Dummett, M. 1979. Common sense and physics. In {\em Perception and identity: essays %%@
presented to A. J. Ayer}, ed. G. Macdonald, London. Reprinted in Dummett (1978).

Dummett, M. 1986. Causal loops. In {\em The nature of time}, ed. R. Flood and M. Lockwood. %%@
Oxford: Oxford University Press. Reprinted in Dummett (1993); page reference to reprint.

Dummett, M. 1993. {\em The seas of language}. Oxford: Oxford University Press.

Dummett, M. 2000. Is time a continuum on instants? {\em Philosophy} 75: 497-515.

Dummett, M. 2003. The Dewey Lectures 2002: Truth and the Past. {\em Journal of Philosophy} %%@
100: 5-53.

Dummett, M. 2004. {\em Truth and the Past}. New York: Columbia University Press.

Dummett, M. 2005. Hume's atomism about events; a response to Ulrich Meyer. {\em Philosophy} %%@
80: 141-144.

Dummett, M. 2005a. The justificationist's response to a realist. {\em Mind} 114: 671-688.

Dummett, M. 2007. The place of philosophy in European culture. {\em European Journal of %%@
Analytic  Philosophy} 3: 21-30.

Earman, J., Smeenk, C. and Wuthrich, C.  2009. Do the laws of physics forbid the operation %%@
of time machines? {\em Synthese} 169: 91-124.

Everett, H. 1957. ``Relative-state'' formulation of quantum mechanics. {\em Reviews of %%@
Modern Physics}, 29: 454-462.

Gale, R. 1964. Is it now now?  {\em Mind} 73: 97-105.

Greaves, H. 2004. Understanding Deutsch's probability in a deterministic multiverse. {\em %%@
Studies in the History and Philosophy of Modern Physics} 35: 423-456.

Kastner, R. 2008. The transactional interpretation, counterfactuals, and weak values in %%@
quantum theory. {\em Studies in the History and Philosophy of Modern Physics} 39: 806-818.

Lewis, D. 1976. The paradoxes of time travel. {\em American Philosophical Quarterly} 13: %%@
145-152. Reprinted in Lewis (1986a); page reference to reprint.

Lewis, D. 1979. Attitudes De Dicto and De Se. {\em Philosophical Review} 88: 513-543. %%@
Reprinted in Lewis (1983); page reference to reprint.

Lewis, D. 1983. {\em Philosophical papers volume I}. Oxford: Oxford University Press.

Lewis, D. 1973. {\em Counterfactuals}. Oxford: Blackwell.

Lewis, D. 1986.  {\em On the plurality of worlds}. Oxford: Blackwell.

Lewis, D. 1986a. {\em  Philosophical papers volume II}. Oxford: Oxford University Press.

Lewis, D. 1980. A Subjectivist's Guide to Objective Chance. In {\em  Studies in inductive %%@
logic and probability}, ed. R. Jeffrey. Berkeley: University of California Press. Reprinted %%@
in Lewis (1986a).

McTaggart, J. 1908. The unreality of time. {\em Mind} 18: 457-484.

McTaggart, J. 1927. {\em The nature of existence}, volume 2, Cambridge: Cambridge %%@
University Press.

Markosian, N. 2004. A defence of presentism. In {\em Oxford studies in metaphysics}, ed. D. %%@
Zimmermann, 47-82. Oxford: Oxford University Press, .

Maudlin, T. 2007. On the passing of time. In his {\em The metaphysics within physics}, %%@
104-142. Oxford: Oxford University Press.

Mellor, D. 1981. {\em Real time}, Cambridge: Cambridge University Press.

Meyer, U. 2005. Dummett on the time continuum, {\em Philosophy} 80: 135-140.

Moretti, L. 2008. Dummett and the problem of the vanishing past. {\em Linguistic and %%@
Philosophical Investigations} 7: 37-47.

Norton, J. 1999. A Quantum Mechanical Supertask. {\em Foundations of Physics} 29: %%@
1265–1302.

Norton, J. 2008. The dome: an unexpectedly simple failure of determinism. {\em Philosophy %%@
of Science} 75: 786-798.

Peacocke, C. 2005. Justification, Realism and the Past. {\em Mind}, 114: 639-70.

Perez Laraudogoitia, J. 2009. Supertasks. {\em Stanford Encyclopedia of Philosophy} \\ 
http://plato.stanford.edu/entries/spacetime-supertasks/

Perry, J. 1979. The Problem of the Essential Indexical. {\em Nous} 13: 3-21.

Price, H. 2008. Toy models for retrocausality. {\em Studies in the History and Philosophy %%@
of Modern Physics} 39: 752-761.

Price, H. 2011. On the flow of time. In {\em The Oxford handbook of philosophy of time}, %%@
ed. C. Callender, 279-311. Oxford: Oxford University Press. 

Prior, A. 1959. Thank goodness that's over. {\em Philosophy} 34: 12-17.

Prior, A. 1970. The notion of the present.  {\em Studium Generale} 23: 245-248. Reprinted  %%@
in {\em The Study of Time}, ed. J.T. Fraser, Berlin: Springer.

Saunders, S., Barrett, J., Kent, A., and D. Wallace eds. 2010.
{\em Many worlds? Everett, quantum theory and reality}, Oxford: Oxford University Press.

Sider, T. 2001. {\em Four-dimensionalism: an ontology of persistence and time}, Oxford: %%@
Oxford University Press.

Smeenk, C. and Wuthrich, C.  2011. Time travel and time machines. In {\em The Oxford %%@
handbook of philosophy of time}, ed. C. Callender, 577-630. Oxford: Oxford University %%@
Press.

Stairs, A. 2006. Kriske, Tupman and quantum logic. In {\em Physical theory and its %%@
interpretation}, ed. W. Demopoulos and I. Pitowsky, 253-272. Dordrecht: Springer.

Tappenden, P. 2011. Evidence and uncertainty in Everett's multiverse. {\em British Journal %%@
for the Philosophy of Science} 62: 99-123.

Tooley, M. 1997. {\em Time, tense and causation}. Oxford: Oxford University Press.

Wallace, D. 2005. Language use in a branching universe. Available at %%@
http://philsci-archive.pitt.edu/archive/00002554/

Wallace, D. 2006. Epistemology quantized: circumstances in which we should come to believe %%@
in the Everett interpretation. {\em British Journal for the Philosophy of Science} 57: %%@
655-689.

Wallace, D. 2012. {\em The emergent multiverse: quantum theory according to the Everett %%@
interpretation}. Oxford: Oxford University Press.\\ \\

Address: Trinity College, Cambridge CB2 1TQ, United Kingdom. Email: jb56@cam.ac.uk

\end{document}